\documentclass[a4paper,12pt]{article}
\usepackage[english]{babel}
\usepackage{url}
\usepackage{layaureo}
\usepackage{fullpage}
\usepackage[dvips]{epsfig}

\usepackage{graphicx}
\usepackage{color}
\usepackage{multirow}

\usepackage{amsthm}
\usepackage{amsmath}
\usepackage{amsfonts}
\usepackage{amssymb}
\usepackage{graphicx}
\usepackage{setspace}
\usepackage{natbib}
\usepackage{booktabs}
\usepackage[colorlinks,breaklinks]{hyperref}
\hypersetup{
    colorlinks,%
    citecolor=black,%
    filecolor=black,%
    linkcolor=black,%
    urlcolor=black
}
\usepackage{lscape}
\usepackage{ctable}

\numberwithin{equation}{section}

\newcommand{\F}{\mathcal{F}}

\newcommand{\R}{\mathbb{R}}
\newcommand{\C}{\mathbb{C}}
\renewcommand{\P}{P}
\newcommand{\Q}{Q}

\allowdisplaybreaks

\title{}

\begin{document}

\begin{center}
{\noindent \textbf{\Large Forward-looking portfolio selection with multivariate non-Gaussian models and the Esscher transform}}

\vspace{18pt}

{\noindent Michele Leonardo Bianchi\textsuperscript{a,*}, Gian Luca Tassinari\textsuperscript{b}}\\
\vspace{10pt}
{\noindent \small\textsuperscript{a}\it Regulation and Macroprudential Analysis Directorate,}\\ 
{\it Banca d'Italia, }\\
{\it Via Milano 53, 00184, Rome, Italy}\\
{\it micheleleonardo.bianchi@bancaditalia.it}\\
{\noindent \small\textsuperscript{*} Corresponding author}\\
\vspace{10pt}
{\noindent \small\textsuperscript{b}\it School of Economics, Management and Statistics,\\ Alma Mater Studiorum,}\\
{\it University of Bologna,}\\
{\it Via Zamboni, 33, 40126, Bologna, Italy}\\
{\it gianluca.tassinari2@unibo.it}\\
\vspace{10pt}
This version: \today\\
\vspace{10pt}
\end{center}


\setcounter{page}{1}


\vspace{25pt}

\noindent {\bf Abstract.} In this study we suggest a portfolio selection framework based on option-implied information and multivariate non-Gaussian models.  The proposed models incorporate skewness, kurtosis and more complex dependence structures among stocks log-returns than the simple correlation matrix. The two models considered are a multivariate extension of the normal tempered stable (NTS) model and the generalized hyperbolic (GH) model, respectively, and  the connection between the historical measure $\P$ and the risk-neutral measure $\Q$ is given by the Esscher transform. We consider an estimation method that simultaneously calibrate the time series of  univariate log-returns and the univariate observed volatility smile. To calibrate the models, there is no need of liquid multivariate derivative quotes. The method is applied to fit a 50-dimensional series of stock returns, to evaluate widely known portfolio risk measures and to perform a portfolio selection analysis.

\vspace{25pt}


\vspace{25pt}


\newpage

\section{Introduction}

A portfolio selection problem is by its nature a multi-dimensional problem, and it is usually solved by considering a multivariate normal distribution, mainly because the complexity of a model increases as soon as one moves from a normal to a non-normal distribution framework. Then, if one wants to define a multivariate non-Gaussian option pricing model, the complexity increases further because the asset return dynamics specified under the historical measure $\P$ cannot be directly used to price options, and a proper change of measure between $\P$ and a possible risk-neutral measure $\Q$ has to be specified. Implied volatilities extracted from option prices contain information about the future behavior of asset returns and for this reason they should be considered in portfolio selection (see \cite{demiguel2013improving}). As observed by \cite{cecchetti2013forward}, risk-neutral distributions extracted from option prices reflect market participant expectations, they are inherently forward-looking. Thus, risk-neutral information may provide more accurate estimates of risk factors distribution and related moments, and they can be used to infer market beliefs about economic events of interest, for instance, in central banks, where they are routinely used for monetary policy and financial stability purposes (see \cite{taboga2015option}). 

In this paper we propose a portfolio selection framework based on non-Gaussian models that takes into account option-implied information. The two models considered are a multivariate extension of the normal tempered stable (NTS) model and the generalized hyperbolic (GH) model, respectively. These models incorporate skewness, kurtosis and more complex dependence structures among stocks log-returns than the simple correlation matrix. 

The multivariate normal tempered stable (MNTS) and the multivariate generalized hyperbolic (MGH) distributions are generalizations of the multivariate normal distribution known as multivariate normal mean-variance mixture distributions. These models share much of the structure of the multivariate normal distribution but they allow both asymmetry and heavy tails. This class of models can be also viewed as multivariate time-changed Brownian motions, as observed by \cite{tb2014ijtaf}.  Furthermore, according to \cite{frahm2004generalized}, both the MNTS and the MGH distributions belong to the class of elliptical variance-mean mixture. Elliptical and generalized elliptical heavy tail distributions have been widely studied (see e.g. \cite{kring2009multi} and \cite{dominicy2013inference}). The MNTS has been proposed by \cite{kim2012measuring} and extensively studied by \cite{btf2015riding} and by \cite{fallahgoul2017quanto}. The MGH is a popular choice when deviating from the multivariate normal distribution towards fatter tailed multivariate distributions (see \cite{protassov2004based}, \cite{hu2005calibration}, \cite{embrechts2005quantitative}, \cite{hu2007risk}, and \cite{hu2010qf}). 

In most of the non-Gaussian continuous-time models, discussed in the literature on option pricing, the change of measure from the historical measure $\P$ to the risk-neutral measure $\Q$ (needed to evaluate the prices of options) is not unique. This means that to find a proper change of measure it is necessary to estimate the model by also considering the prices of options traded in the market. In this study the connection between the historical measure $\P$ and the risk-neutral measure $\Q$ is given by the Esscher transform (see \cite{gerber1995option}, \cite{sato1999levy}, and \cite{tb2014ijtaf}).  We discuss a possible approach to jointly estimate historical asset returns (under the historical measure) and calibrate option implied volatilities (under the risk-neutral measure). \cite{tb2014ijtaf} proposed a joint calibration-estimation of the univariate option surfaces and of the time series of log-returns by considering the EM-based maximum likelihood estimation method together with the multivariate Esscher transform needed to find the link between historical and risk-neutral parameters. The authors have referred to this joint calibration-estimation as {\it double calibration}. In this paper we apply the same approach. A similar calibration procedure which rests on a joint calibration of univariate option surfaces and pairwise correlations has been introduced by \cite{guillaume2012sato} and \cite{guillaume2012alphaVG}. Recently, \cite{ballotta2017multivariate} proposed a joint calibration to market quotes for both FX triangles and quanto products.


The risk measure we use in this study is the average value at risk (AVaR), the average of the values-at-risk (VaRs) greater than the VaR at a given tail probability. AVaR, also called conditional value-at-risk (CVaR) or expected shortfall, is a superior risk measure to VaR because it satisfies all axioms of a coherent risk measure and it is consistent with preference relations of risk-averse investors (see \cite{rsf2008}). Thus, we follow a mean-AVaR portfolio selection criterion which by construction, not only takes into consideration the first two moments of the distribution but also the behavior in its left tail. For the models we proposed the AVaR has a closed formula (up to a numerical integration), which can be easily cast into a portfolio optimization problem. In the empirical analysis we consider minimum-AVaR (MA) portfolios (see \cite{stoyanov2010survey}) under the normal, MGH and MNTS distributional assumption and, as done in similar studies (e.g. \cite{demiguel2007optimal} 
and \cite{mainik2015portfolio}), we compare them with two benchmark portfolios, that is the minimum-variance portfolio (MV) and the equally weighed portfolio (EW).

The paper is organized as follows. In Sections \ref{sec:MNTS} and \ref{sec:MGH} we define the multivariate normal mean-variance mixture distributions with tempered stable and generalized inverse Gaussian mixing distribution, respectively. Then, we explain how to find the risk-neutral parameters starting from the historical ones by means of the multivariate Esscher transform. Additionally, we show how it is possible to find the historical parameters starting from the risk-neutral ones. In Section \ref{sec:Calibration} we briefly describe the data analyzed in the empirical study and in Section \ref{sec:TripleCalibration} we describe the double calibration algorithm in which we simultaneously calibrate the implied volatility surface, by minimizing the average relative percentage error (a measure of the distance between model and observed implied volatilities), and estimate the model parameters on the time series of log-returns by simultaneously minimizing the Kolmogorov-Smirnov distance. 
In Section \ref{sec:Portfolio} we review the minimum-AVaR portfolio selection approach, we describe the main computational aspects, and we discuss the results. After having resumed the main results, Section \ref{sec:Conclusions} concludes.


\section{Multivariate option pricing models}\label{sec:Model}

In this section we review the model described in \cite{tb2014ijtaf}. We analyze a market with $n$ stocks, and we assume that the dynamics of asset log-returns are described through a L\'evy process obtained by a multivariate Brownian motion, time-changed by a single one-dimensional non-negative non-decreasing L\'evy process. We refer to this last L\'evy process as a {\it subordinator} or a {\it stochastic clock}. We further assume that the components of the multivariate Brownian motion are correlated, and the subordinator is independent of the Brownian motion. Building a process in this way allows to get two sources of dependence: (1) a jump in the subordinator  produces a jump in the price processes, and all jumps occur at the same time; furthermore, (2) since we assume a correlated Brownian motion, the jump sizes are correlated. 

The price at time $t$ of the stock $j$ is given by the following equation
\begin{equation}
	P_t^{j}=P_0^{j}\exp(Y_t^{j})\label{eq:levyprices},	
\end{equation}	  
where $P_t^{j}$ and $P_0^{j}$ are the price of the stock $j$ at times $t$ and $0$, respectively, and $Y_t^{j}$ is the log-return of the $j$-th underlying asset over the interval $\left[0, t\right]$, for every $j=1,\ldots, n$. Thus, the log-return process of the $j$-th underlying asset $Y^j=\{Y_t^j,\ t\geq0\}$ is defined as
\begin{equation}\label{eq:SubordinatedMarginals}
Y_t^{j}=\mu_j t +X_t^{j}=\mu_jt+\theta_j S_t+\sigma_j W^{j}_{S_t},	
\end{equation}
where $X^{j}=\{X^{j}_t,$ $t\geq0\}$ is the pure jump part of the process, $S=\{S_t,\ t\geq0\}$ is the subordinator, $W^{j}=\{W^{j}_t,$ $t\geq0\}$ and $W^{k}=\{W^{k}_t,\ t\geq0\}$ are correlated one-dimensional Brownian motions with correlation coefficient $\rho_{jk}$, $W_{S}^{j}=\{W_{S_t}^j,\ t\geq0\}$ is the $j$-th Brownian motion evaluated at the common stochastic clock $S_t$. Then, $\mu$, $\theta$ and $\sigma$ are vectors in $\R^n$, and the elements of the vector $\sigma$ are strictly positive ($\sigma_j>0$). As shown in \cite{tb2014ijtaf},  there exists a relation between subordinated multivariate Brownian motions and  multivariate distributions defined as normal mean-variance mixture and this relation is useful to implement the expectation-maximization (EM) maximum likelihood estimation (see \cite{embrechts2005quantitative}).

The characteristic function of the multivariate pure jump part process $X=\{X_t,\ t\geq0\}$ can be computed by considering equation (4.6) in \cite{cont2003financial}, that is
\begin{equation}\label{eq:cfPUREjump}
\Psi_{X_t}\left(u\right)=\exp\left(tl_{S_1}(g(u))\right),
\end{equation}
where $l_{S_1}\left(.\right)$ is the Laplace exponent of the common subordinator and $g\left(u\right)$ is the characteristic exponent of the multivariate Brownian motion, that is
\begin{equation}\label{eq:ChExpMBm1}
\begin{split}
g\left(u\right)&\;= i u'\theta-\frac{1}{2} u'\Sigma u\\  
&\;= \sum_{j=1}^n i u_j\theta_{j}-\frac{1}{2}\sum_{j=1}^n \sum_{k=1}^n u_{j} u_k \sigma_{j}\sigma_{k}\rho_{jk},
\end{split}
\end{equation}
with $u \in {\R}^n$, and where the matrix $\Sigma$ has elements $\Sigma_{jk} = \sigma_{j}\sigma_{k}\rho_{jk}$. Since $\Sigma$ is a variance-covariance matrix, we can rewrite equation (\ref{eq:ChExpMBm1}) in the following form 
\begin{equation}\label{eq:ChExpMBm}
g\left(u\right)= i u'\theta-\frac{1}{2} u'D_{\sigma}\Omega D_{\sigma} u,
\end{equation}
where $D_{\sigma}$ is a diagonal matrix with diagonal $\sigma\in\R^n$, and $\Omega$ is the correlation matrix of the Brownian motions with elements $\rho_{jk}$. The characteristic function of the log-return process $Y=\{Y_t,\ t\geq0\}$ is given by
\begin{equation}\label{eq:cf}
 \Psi_{Y_t}\left(u\right)=\exp\left(itu'\mu\right)\Psi_{X_t}\left(u\right).
\end{equation}

Then, assuming the existence of a bank account which provides a continuously compounded risk-free rate $r$ constant over the interval $[0, t]$, this market model is arbitrage free, since the price process of every asset has both positive and negative jumps (see \cite{cont2003financial}). This ensures the existence of an equivalent martingale measure. However, the model is not complete, and, therefore, the risk-neutral measure is not unique. Among the possible candidates we select the Esscher equivalent martingale measure (EEMM) proposed in \cite{gerber1995option}. The $Q_h$ Esscher measure associated with the multivariate log-return process $Y_t$ is defined by the following Radon-Nikodym derivative
\begin{equation}
	\frac{dQ_h}{dP}|\F_t=\frac{\exp(h' Y_t)}{E\left[\exp(h'Y_t)\right]}\label{eq:ERND},
\end{equation}
where $\F_t$ is the filtration originated by the price processes. 

To get the Esscher risk-neutral dynamics of $Y_t$ we need to find a vector $h$ such that the discounted price process of every asset is a martingale under the new probability measure $Q_h$, that is,
\begin{equation}\label{eq:EsscherSystemInitial}
\begin{cases}
E^{Q_h}\left[P_t^{1}\exp\left[(-r+d_1)t\right]\right]=E\left[P_t^{1}\exp\left[(-r+d_1)t\right]\frac{\exp(h' Y_t)}{E\left[\exp(h'Y_t)\right]}\right]=P_0^{1},\\ 
  \vdots&\\
E^{Q_h}\left[P_t^{n}\exp\left[(-r+d_n)t\right]\right]=E\left[P_t^{n}\exp\left[(-r+d_n)t\right]\frac{\exp(h' Y_t)}{E\left[\exp(h'Y_t)\right]}\right]=P_0^{n},
\end{cases}
\end{equation}
where $d_1,...,d_n$ are the continuously compounded dividend yields of the $n$ stocks. 
Substituting (\ref{eq:levyprices}) in (\ref{eq:EsscherSystemInitial}), after some computations we obtain
\begin{equation}\label{eq:EsscherSystem}
\begin{cases}
E\left[\exp(h' Y_t+Y_t^1)\right]/E\left[\exp(h' Y_t)\right]=\exp\left[(r-d_1)t\right],\\ 
  \vdots&\\
E\left[\exp(h' Y_t+Y_t^n)\right]/E\left[\exp(h' Y_t)\right]=\exp\left[(r-d_n)t\right].  
\end{cases}
\end{equation}
We cannot be sure that an equivalent martingale measure exists. However, if it is possible to find a vector $h$
such that the discounted price process of each stock is a martingale under the measure $Q_h$, then the existence of the EEMM is ensured.
Assuming that a solution exists, it is possible to compute the characteristic function of the process $Y_t$ under the probability measure $Q_h$, that is
\begin{equation} 
\Psi^{Q_h}_{Y_t}\left(u\right)=E\left[\exp((iu+h)' Y_t)\right]/E\left[\exp(h' Y_t)\right].
\end{equation}
Since the equality  
$$
\Psi_{Y_t}\left(u\right)=E\left[\exp(iu' Y_t)\right]
$$
holds by definition of the characteristic function, it is possible to prove that
\begin{equation}\label{eq:CfRn} 
\Psi^{Q_h}_{Y_t}\left(u\right)=\Psi_{Y_t}\left(u-ih\right)/\Psi_{Y_t}\left(-ih\right).
\end{equation}
Given a vector $h$, equation (\ref{eq:CfRn}) shows that the characteristic function under the measure $Q_h$ can be expressed as a function of the characteristic function under the measure $P$. By definition, the characteristic function is defined on $\R$. When we evaluate it on non-real values, such as in equation (\ref{eq:CfRn}), we consider the analytical extension of the characteristic function on a strip of the complex plane $\C$ (eventually, the whole complex plane, if the characteristic function is entire).

Given a set of parameters under the historical measure $P$, one can find a vector $h$ satisfying the equalities (\ref{eq:EsscherSystem}) and the corresponding parameters under the risk-neutral measure $Q_h$. Similarly, given the parameters under the risk-neutral measure $Q$, the inverse procedure can be implemented to find the parameters under the historical measure $P_h$. We refer to this measure change as the {\it inverse Esscher transform}. The inverse Esscher transform will allow us to introduce the calibration procedure proposed in Section \ref{sec:TripleCalibration}. More specifically, in order to find the inverse Esscher transform, that is the vector $h$ that allows one to find the historical parameters (under the historical measure $P_h$) starting from the risk-neutral ones (under the risk-neutral measure $Q$), we have to solve the following system 
\begin{equation} \label{eq:InvEsscherSystem}
\begin{cases}
E^{Q}\left[\exp(Y_t^1)\right]=\exp\left[(r-d_1)t\right],\\ 
  \vdots&\\
E^{Q}\left[\exp(Y_t^n)\right]=\exp\left[(r-d_n)t\right],  
\end{cases}
\end{equation}
under some constraints imposed by some functional relations between historical and risk-neutral parameters that will be discussed in the sequel.

\subsection{The multivariate NTS model}\label{sec:MNTS}

A formal elegant definition of tempered stable (TS) distributions and processes was proposed in the work of \cite{rosinski2007tempering} and applied to finance in numerous empirical studies (see \cite{rachev2011financial}), principally under a univariate framework (see \cite{bianchi2014open} and references therein). In this section we study the multivariate extension of the normal tempered stable model proposed by \cite{kim2012measuring} and extensively studied by \cite{btf2015riding}. Let $S=\{S_t,\ t\geq0\}$ be a classical tempered stable process, that is, a process which starts at zero and has stationary and independent increments, in which the law of $S_1$ is classical tempered stable (CTS) with parameters $\lambda> 0$, $C> 0$ and $0<\omega<2$. We refer to the law $S_1$ as $CTS(\omega, \lambda, C)$. The characteristic function of the random variable $S_1$ is
\begin{equation}\label{eq:CfTS}
\Psi_{S_1}{(u)}=\exp\left( C\Gamma(-\omega)((\lambda - iu)^{\omega} -\lambda^{\omega}))\right).
\end{equation}


From (\ref{eq:CfTS}) it is possible to compute the Laplace exponent of the classical tempered stable subordinator
\begin{equation} 
l_{S_1}(u)=\ln\Psi_{I_1}{\left(\frac{u}{i}\right)}=C\Gamma(-\omega)((\lambda - u)^{\omega} -\lambda^{\omega})
\end{equation}
and the moments of $S_1$
\begin{equation}
	E\left[S_{1}\right]=-\omega C\Gamma(-\omega) \lambda^{\omega -1},
\end{equation}	
\begin{equation}		
var\left[S_{1}\right]=\omega (\omega -1) C \Gamma(-\omega) \lambda^{\omega -2},
\end{equation}
\begin{equation}	
skew\left[S_{1}\right]=(2-\omega) \left[\omega (\omega -1) C \Gamma(-\omega) \lambda^{\omega}\right]^{-\frac{1}{2}},
\end{equation}
\begin{equation}		
kurt\left[S_{1}\right]=3+(\omega -2)(\omega -3)\left[\omega (\omega -1) C \Gamma(-\omega) \lambda^{\omega}\right]^{-1}.
\end{equation}

Finally, using (\ref{eq:cf}) we get the characteristic function of the multivariate normal classical tempered stable (MNTS) process with linear drift
\begin{equation}\label{eq:ChfMNTS}
\Psi_{Y_t}\left(u\right)=\exp\left\{t \left[ iu'\mu + C\Gamma\left(-{\omega}\right)\left(\left(\lambda - iu'\theta + \frac{1}{2} u'\Sigma u\right)^{\omega} - \lambda^{\omega} \right)\right]\right\}.
\end{equation}
Setting $u_i=0$, $\forall i\neq j$, into (\ref{eq:ChfMNTS}) we get the characteristic function of the log-return process of the $j$-th underlying asset
\begin{equation}\label{eq:CfMTSMar} 
\Psi_{Y_{t}^j}\left(u_j\right)=\exp\left\{t \left[ iu_j\mu_j + C\Gamma\left(-{\omega}\right)\left(\left(\lambda - iu_j\theta_j + \frac{1}{2} u_j^2\sigma_j^2\right)^{\omega} - \lambda^{\omega} \right)\right]\right\}.
\end{equation}
If we set $\omega=1/2$, $S_1$ follows an inverse Gaussian distribution with parameters $\gamma=-\frac{C\Gamma(-\omega)}{\sqrt{2}}$ and $\eta=\sqrt{2 \lambda}$.
If $\omega\rightarrow 0$, $S_1$ follows a gamma distribution with parameters $\alpha=-\frac{C\Gamma(-\omega)}{\sqrt{2}}$ and $\beta=\sqrt{2 \lambda}$. In the first case our MNTS model leads to the multivariate normal inverse Gaussian (MNIG) and in the second case to the multivariate variance gamma (MVG) of \cite{tb2014ijtaf}.

From (\ref{eq:ChfMNTS}) it is possible to compute marginal and joint moments of log-return increments over the period $\left[0, t\right]$ and express them as functions of the moments of the subordinator 
\begin{equation}
E\left[Y_{t}^j\right]=\mu_j t +	E\left[S_{t}\right]\theta_j,
\end{equation}
\begin{equation}		
var\left[Y_{t}^j\right]=var\left[S_{t}\right]\left(\theta_j^2+ \frac{ \sigma_{j}^2 \lambda}{1-\omega}\right),
\end{equation}
\begin{equation}	
skew\left[Y_{t}^j\right]=skew\left[S_{t}\right]\left(\theta_j^3+ \frac{3\theta_j \sigma_{j}^2 \lambda}{2-\omega}\right)\left(\theta_j^2+ \frac{ \sigma_{j}^2 \lambda}{1-\omega}\right)^{-\frac{3}{2}},
\end{equation}
\begin{equation}		
kurt\left[Y_{t}^j\right]=3+\left(kurt\left[S_{t}\right]-3\right)\left[\theta_j^4+ \frac{3\sigma_{j}^2 \lambda}{3-\omega}\left(2 \theta_j^2+ \frac{\sigma_{j}^2 \lambda}{2-\omega}\right)\right]\left(\theta_j^2+ \frac{ \sigma_{j}^2 \lambda}{1-\omega}\right)^{-2},
\end{equation}
\begin{equation}	
cov\left[Y_{t}^i; Y_{t}^j\right]=var\left[S_{t}\right]\left(\theta_i\theta_j+ \frac{ \sigma_{ij} \lambda}{1-\omega}\right),
\end{equation}
\begin{equation} \label{eq:CorrelationMNTS}
corr\left[Y_{t}^i; Y_{t}^j\right]=\frac{\theta_i\theta_j+ \frac{ \sigma_{ij} \lambda}{1-\omega}}{\sqrt{\left(\theta_i^2+ \frac{\sigma_{i}^2 \lambda}{1-\omega}\right)\left(\theta_j^2+ \frac{ \sigma_{j}^2 \lambda}{1-\omega}\right)}}.
\end{equation}


Then, in order to find the Esscher risk-neutral dynamics of the log-return process, we follow the procedure described in Section \ref{sec:Model}. In the rest of the paper, if not differently stated, the parameters are under the historical measure $P$. More precisely, by considering equations (\ref{eq:ChfMNTS}) and (\ref{eq:EsscherSystem}), the system to solve to find the vector $h$ may be written as
\begin{equation}\label{eq:MNTSSystem}
\begin{cases}
\left(\lambda-l\left(h\right)\right)^{\omega}-\left(\lambda-\theta_1-\frac{1}{2}\sigma_1^2-l\left(h\right)-\sum_{j=1}^{n}h_j\sigma_{1}\sigma_{j}\rho_{1j}\right)^{\omega}&=(\mu_1+d_1-r)/C\Gamma(-\omega),\\ 
  &\vdots\\
\left(\lambda-l\left(h\right)\right)^{\omega}-\left(\lambda-\theta_n-\frac{1}{2}\sigma_n^2-l\left(h\right)-\sum_{j=1}^{n}h_n\sigma_{n}\sigma_{j}\rho_{nj}\right)^{\omega}&=(\mu_n+d_n-r)/C\Gamma(-\omega),\\ 
\end{cases}
\end{equation}
where
\begin{equation}
l\left(h\right)=h'\theta+\frac{1}{2} h'\Sigma h.
\end{equation}
Although the existence of a solution for the system (\ref{eq:MNTSSystem}) is difficult to prove, in all of our applications it is possible to solve the system numerically. This ensures that at least an EEMM exists.

Using (\ref{eq:CfRn}) we get the Esscher risk-neutral characteristic function
\begin{equation}\label{eq:ChfMNTSRn}
\Psi_{Y_t}\left(u\right)=\exp\left\{t \left[ iu'\mu^{Q_h} + C^{Q_h}\Gamma\left(-{\omega^{Q_h}}\right)\left(\left(\lambda^{Q_h} - iu'\theta^{Q_h} + \frac{1}{2} u'\Sigma^{Q_h} u\right)^{\omega^{Q_h}} - {\lambda^{Q_h}}^{\omega^{Q_h}} \right)\right]\right\}.
\end{equation}
where the relations among risk-neutral and historical parameters are
\begin{equation}\label{eq:MNTSEsscherTransformParameters}
\begin{split}
\mu^{Q_h}&\;=\mu,\\
C^{Q_h}&\;=C,\\
\omega^{Q_h}&\;=\omega,\\	
\lambda^{Q_h}&\;=\lambda-h'\theta-\frac{1}{2} h'\Sigma h,\\	
\theta^{Q_h}&\;=\theta + \Sigma h,\\	
D_{\sigma^{Q_h}}&\;=D_{\sigma},\\
\Omega^{Q_h}&\;=\Omega.\\	
\end{split}
\end{equation}
The expression of the Esscher risk-neutral characteristic function of the $j$-th log-return process is
\begin{equation}\label{eq:CfMNTSMarginalRn}
\Psi_{Y_t^{j}}\left(u\right)=\exp\left\{t \left[ iu'\mu_j^{Q_h} + C^{Q_h}\Gamma\left(-{\omega^{Q_h}}\right)\left(\left(\lambda^{Q_h} - iu'\theta_j^{Q_h} + \frac{1}{2} u_j^2 {\sigma_j^{Q_h}}^{2}\right)^{\omega^{Q_h}} - {\lambda^{Q_h}}^{\omega^{Q_h}} \right)\right]\right\}.
\end{equation}

Note that the Esscher change of probability measure does not modify the nature of joint and marginal log-return processes (compare (\ref{eq:ChfMNTS}) with (\ref{eq:ChfMNTSRn})),  
and (\ref{eq:CfMTSMar}) with (\ref{eq:CfMNTSMarginalRn})). Only the parameter $\lambda$ and the vector $\theta$ change. In particular, the risk-neutral log-return process is a multivariate Brownian motion with correlated components, time-changed by a single classical tempered stable subordinator independent of the Brownian motion. Even if the correlation matrix of the underlying Brownian motion is not affected by the change of measure, the risk-neutral correlation matrix of log-returns is different (see (\ref{eq:MNTSEsscherTransformParameters}) and (\ref{eq:CorrelationMNTS})). Notice also that all marginal moments change (not only the mean). It should also be pointed out that all risk-neutral moments depend on the entire dependence structure.  As shown in equation (\ref{eq:MNTSEsscherTransformParameters}), 
the risk-neutral distribution depends on the correlation matrix of the underlying Brownian motions under the historical measure.

In order to find the inverse Esscher transform, that is the vector $h$ that allows to find the historical parameters (under the historical measure $P_h$) starting from the risk-neutral ones (under the risk-neutral measure $Q$), from the system (\ref{eq:InvEsscherSystem}) it follows that the system to be solved to find $h$ is the following
\begin{equation}\label{eq:InverseMNTSSystem}
\begin{cases}
\mu_1^{Q} + C^{Q}\Gamma\left(-{\omega^{Q}}\right)\left(\left(\lambda^{Q} - \theta_1^{Q} - \frac{1}{2} {\sigma_1^{Q}}^{2}\right)^{\omega^{Q}} - {\lambda^{Q}}^{\omega^{Q}} \right)=r-d_1,\\ 
  \vdots&\\
\mu_n^{Q} + C^{Q}\Gamma\left(-{\omega^{Q}}\right)\left(\left(\lambda^{Q} - \theta_n^{Q} - \frac{1}{2} {\sigma_n^{Q}}^{2}\right)^{\omega^{Q}} - {\lambda^{Q}}^{\omega^{Q}} \right)=r-d_n,\\ 
\mu^{P_ h}=\mu^{Q},\\
C^{P_ h}=C^{Q},\\	
\omega^{P_ h}=\omega^{Q},\\	
\lambda^{P_ h}={\lambda^{Q}}+h'\theta^Q-\frac{1}{2} h'\Sigma^{Q} h,\\	
\theta^{P_ h}=\theta^{Q} - \Sigma^{Q} h,\\	
D_{\sigma^{P_h}}=D_{\sigma^{Q}},\\
\Omega^{P_h}=\Omega^{Q}.\\	
\end{cases}
\end{equation}
In solving this system one has to consider that, given the parameters under the measure $Q$, both the vector $h$ and the parameters under the measure $P_h$ are the system unknowns to be found. As discussed in Section \ref{sec:Model}, the system has to be solved under the constraints in equation (\ref{eq:MNTSEsscherTransformParameters}) and, thus, the system unknowns are the vector $h$ together with the parameters $\lambda^{P_ h}$ and $\theta^{P_ h}$.

\subsection{The multivariate GH model}\label{sec:MGH}

The generalize hyperbolic (GH) distribution has received a lot of attention in the financial-modeling literature (see for example \cite{eberlein1995hyperbolic}, \cite{prause1999generalized}, and \cite{eberlein2002generalized}). The GH parametric family includes many familiar distributions such as for example the Student's $t$, the skew-$t$, the hyperbolic, the variance gamma, and the normal inverse Gaussian. In this section we study the multivariate generalized hyperbolic (MGH) distribution. Let $G=\{G_t,\ t\geq0\}$ be a generalized inverse Gaussian process (GIG), i.e., a process which starts at zero and has stationary and independent increments, in which the distribution of $G_1$ is generalized inverse Gaussian with parameters $\epsilon$, $\psi$, $\chi$.  $\psi$ and $\chi$ are are both nonnegative and not simultaneously 0. We refer to the law of $G_1$ as GIG $\left(\epsilon, \chi, \psi \right)$ 
The density function of the random variable $G_1$ is
\begin{equation}\label{eq:DensityGIG}
f(x;\epsilon, \psi, \chi)= \frac{1}{2K_{\epsilon}\left(\sqrt{\chi\psi}\right)}\left(\frac{\psi}{\chi}\right)^{\frac{\epsilon}{2}} x^{\epsilon-1}\exp\left[-\frac{1}{2}\left(\frac{\chi}{x} + \psi x\right)\right], x>0,  
\end{equation}
and its characteristic function is
\begin{equation}\label{eq:CfGIG}
\Psi_{G_1}{(u)}=\left(1-\frac{2iu}{\psi}\right)^{-\frac{\epsilon}{2}}\frac{K_{\epsilon}\left(\sqrt{\chi(\psi-2i u)}\right)}{K_{\epsilon}\left(\sqrt{\chi\psi}\right)}.
\end{equation}

From (\ref{eq:CfGIG}) it is possible to compute the Laplace exponent of the generalized inverse Gaussian subordinator
\begin{equation}\label{eq:LaplaceGIG}
l(u)=\ln\Psi_{G_1}{\left(\frac{u}{i}\right)}=-\frac{\epsilon}{2}\ln\left(1-\frac{2u}{\psi}\right)+\ln\frac{K_{\epsilon}\left(\sqrt{\chi(\psi-2i u)}\right)}{K_{\epsilon}\left(\sqrt{\chi\psi}\right)}
\end{equation}
and from equation (\ref{eq:LaplaceGIG}) the cumulants of $G_1$
\begin{equation}
c_1\left[G_{1}\right] = E\left[G_{1}\right]=\left(\frac{\chi}{\psi}\right)^\frac{1}{2}\frac{ K_{\epsilon+1}\left(\sqrt{\chi\psi}\right) }{ K_{\epsilon}\left(\sqrt{\chi\psi}\right)},
\end{equation}
\begin{equation}		
c_2\left[G_{1}\right] = var\left[G_{1}\right]=\left(\frac{\chi}{\psi}\right)\left[\frac{K_{\epsilon+2}\left(\sqrt{\chi\psi}\right)}{K_{\epsilon}\left(\sqrt{\chi\psi}\right)}-\left(\frac{K_{\epsilon+1}\left(\sqrt{\chi\psi}\right)}{K_{\epsilon}\left(\sqrt{\chi\psi}\right)}\right)^2\right],
\end{equation}
\begin{equation}
c_3\left[G_{1}\right] =\left(\frac{\chi}{\psi}\right)^\frac{3}{2}\left[\frac{K_{\epsilon+3}\left(\sqrt{\chi\psi}\right)}{K_{\epsilon}\left(\sqrt{\chi\psi}\right)}-\frac{3K_{\epsilon+2}\left(\sqrt{\chi\psi}\right)K_{\epsilon+1}\left(\sqrt{\chi\psi}\right)}{K^2_{\epsilon}\left(\sqrt{\chi\psi}\right)}+2\left(\frac{K_{\epsilon+1}\left(\sqrt{\chi\psi}\right)}{K_{\epsilon}\left(\sqrt{\chi\psi}\right)}\right)^3\right],
\end{equation}	
\begin{equation}
\begin{split}		
c_4\left[G_{1}\right]&= \left(\frac{\chi}{\psi}\right)^2\left[\frac{K_{\epsilon+4}\left(\sqrt{\chi\psi}\right)}{K_{\epsilon}\left(\sqrt{\chi\psi}\right)}-\frac{4K_{\epsilon+3}\left(\sqrt{\chi\psi}\right)K_{\epsilon+1}\left(\sqrt{\chi\psi}\right)}{K^2_{\epsilon}\left(\sqrt{\chi\psi}\right)}-3\left(\frac{K_{\epsilon+2}\left(\sqrt{\chi\psi}\right)}{K_{\epsilon}\left(\sqrt{\chi\psi}\right)}\right)^2 \right]+\\&+6\left(\frac{\chi}{\psi}\right)^2\left[\frac{2 K_{\epsilon+2}\left(\sqrt{\chi\psi}\right)K^2_{\epsilon+1}\left(\sqrt{\chi\psi}\right)}{K^3_{\epsilon}\left(\sqrt{\chi\psi}\right)}-\left(\frac{K_{\epsilon+1}\left(\sqrt{\chi\psi}\right)}{K_{\epsilon}\left(\sqrt{\chi\psi}\right)}\right)^4 \right].
\end{split}
\end{equation}

Finally, using (\ref{eq:cf}) we get the characteristic function of the multivariate generalized hyperbolic (MGH) process with linear drift 
\begin{equation}\label{eq:ChfMGH}
\begin{split} 
\Psi_{Y_t}\left(u\right)&=\exp\left(i u' \mu t\right)\left[1-\frac{2}{\psi}\left(iu'\theta-\frac{1}{2} u'\Sigma u\right)\right]^{-\frac{\epsilon t}{2}} \left( \frac{K_{\epsilon}\left(\sqrt{\chi\left(\psi-2\left(iu'\theta-\frac{1}{2} u'\Sigma u\right)\right)}\right)}{K_{\epsilon}\left(\sqrt{\chi\psi} \right)}\right)^t
\end{split} 
\end{equation}
Setting $u_i=0$, $\forall i\neq j$, into (\ref{eq:ChfMGH}) we get the characteristic function of the log-return process of the $j$-th underlying asset
\begin{equation}\label{eq:CfMGHMar} 
\Psi_{Y_{t}^j}\left(u_j \right)=\exp\left(i u_j \mu_j t\right)\left[1-\frac{2}{\psi}\left(iu_j\theta_j-\frac{1}{2} u_j^2\sigma_j^2\right)\right]^{-\frac{\epsilon t}{2}} \left(  \frac{ K_{\epsilon} \left(\sqrt{\chi\left(\psi-2\left(iu_j\theta_j-\frac{1}{2}u_j^2\sigma_j^2\right)\right)}\right)}{K_{\epsilon}\left(\sqrt{\chi\psi} \right)}\right) ^t.
\end{equation}

If we set $\epsilon=-1/2$, $G_1$ follows an inverse Gaussian distribution with parameters $\gamma=\sqrt{\chi}$ and $\eta=\sqrt{\psi}$.
If we set $\chi=0$, $G_1$ follows a gamma distribution $\alpha=\epsilon$ and $\beta=\psi/2$. In the first case we get the MNIG model and in the second one the MVG of \cite{tb2014ijtaf}.
 
From (\ref{eq:ChfMGH}) it is possible to compute marginal and joint moments of log-return increments over the period $\left[0, t\right]$ and express them as functions of the moments and cumulants of the subordinator of log-returns distribution over the period $\left[0; t\right]$: 
\begin{equation}
E\left[Y_{t}^j\right]=\mu_j t + E\left[G_{t}\right]\theta_{j}
\end{equation}
\begin{equation}		
var\left[Y_{t}^j\right]=E\left[G_{t}\right]\sigma_j^2+var\left[G_{t}\right] \theta_j^2
\end{equation}	
\begin{equation}	
skew\left[Y_{t}^j\right]=3 var\left[G_{t}\right]\theta_j\sigma_j^2+  c_3\left[G_{t}\right]\theta_j^3
\end{equation}
\begin{equation}		
ex.kurt\left[Y_{t}^j\right]=3 var\left[G_{t}\right]\sigma_j^4+ 6c_3\left[G_{t}\right]\theta_j\sigma_j^2+c_4\left[G_{t}\right]\theta^4_j
\end{equation}
\begin{equation}		
cov\left[Y_{t}^j; Y_{t}^k\right]=E\left[G_{t}\right]\sigma_{jk}+V\left[G_{t}\right] \theta_j\theta_k
\end{equation}
\begin{equation}	\label{eq:CorrelationMGH}
corr\left[Y_{t}^j; Y_{t}^k\right]=\frac{\sigma_{jk}+ \theta_j\theta_k\Delta\left(\frac{\chi}{\psi}\right)^\frac{1}{2}}{\sqrt{\left[\sigma_{j}^2+ \theta_{j}^2\Delta\left(\frac{\chi}{\psi}\right)^\frac{1}{2}\right]\left[\sigma_{k}^2+ \theta_{k}^2\Delta\left(\frac{\chi}{\psi}\right)^\frac{1}{2}\right]}}
\end{equation}
where
\begin{equation}
	\Delta=\left(\frac{K_{\epsilon +2}\left(\sqrt{\chi\psi}\right)}{K_{\epsilon +1}\left(\sqrt{\chi\psi}\right)}-\frac{K_{\epsilon +1}\left(\sqrt{\chi\psi}\right)}{K_{\epsilon}\left(\sqrt{\chi\psi}\right)}\right).
\end{equation}

%

Then, in order to find the Esscher risk-neutral dynamics of the log-return process, we follow the procedure described in Section \ref{sec:Model}. In the rest of the paper, if not differently stated, the parameters are under the historical measure $P$. More precisely, by considering equations (\ref{eq:ChfMGH}) and (\ref{eq:EsscherSystem}), the system to solve to find the vector $h$ may be written as
\begin{equation} \label{eq:MGHSystem}
\begin{cases}
\left[1-\frac{2\left(\theta_1+0.5\sigma_1^2+\sum_{j=1}^{n}h_j\sigma_{1j}\right)}{\psi-2l\left(h\right)}\right]^{-\frac{\epsilon}{2}} \frac{K_{\epsilon}\left(\sqrt{\chi\left(\psi-2 l\left(h\right)-2\left(\theta_1+0.5\sigma_1^2+\sum_{j=1}^{n}h_j\sigma_{1j}\right)\right)}\right)}{K_{\epsilon}\left( \sqrt{\chi\left(\psi-2 l\left(h\right)\right)}\right)}&=\exp(r-\mu_1-d_1)\\ 
  &\vdots\\
\left[1-\frac{2\left(\theta_n+0.5\sigma_n^2+\sum_{j=1}^{n}h_j\sigma_{nj}\right)}{\psi-2l\left(h\right)}\right]^{-\frac{\epsilon}{2}} \frac{K_{\epsilon}\left(\sqrt{\chi\left( \psi-2 l\left(h\right)-2\left(\theta_n+0.5\sigma_n^2+\sum_{j=1}^{n}h_j\sigma_{nj}\right)\right)}\right)}{K_{\epsilon}\left(\sqrt{\chi\left(\psi-2 l\left(h\right)\right)}\right)}&=\exp(r-\mu_n-d_n)\\ 
\end{cases}
\end{equation}
where
\begin{equation}
l\left(h\right)=h'\theta+\frac{1}{2} h'\Sigma h.
\end{equation}
Although the existence of a solution for the system (\ref{eq:MGHSystem}) is difficult to prove, in all of our applications it is possible to solve the system numerically. This ensures that at least an EEMM exists.

Using (\ref{eq:CfRn}) we get the Esscher risk-neutral characteristic function
\begin{equation} \label{eq:CfMGHRn}
\begin{split} 
\Psi^{Q_h}_{Y_1}\left(u\right)&=\exp\left(i u' \mu^{Q_ h}\right)\left[1-\frac{2}{\psi^{Q_ h}}\left(iu'\theta^{Q_ h}-\frac{1}{2} u'\Sigma^{Q_ h} u\right)\right]^{-\frac{\epsilon^{Q_ h}}{2}} \times\\& \frac{K_{\epsilon^{Q_ h}}\left(\sqrt{\chi^{Q_ h}\left( \psi^{Q_ h}-2\left(iu'\theta^{Q_ h}-\frac{1}{2} u'\Sigma^{Q_ h} u\right)\right)}\right)}{K_{\epsilon^{Q_ h}}\left(\sqrt{\chi^{Q_ h} \psi^{Q_ h} }\right)}
\end{split} 
\end{equation}
where the relations among risk-neutral and historical parameters are
\begin{equation}\label{eq:MGHEsscherTransformParameters}
\begin{split}
\mu^{Q_h}&\;=\mu,\\
\chi^{Q_h}&\;=\chi,\\
\epsilon^{Q_h}&\;=\epsilon,\\	
\psi^{Q_h}&\;=\psi-2\left(h'\theta+\frac{1}{2} h'\Sigma h\right),\\	
\theta^{Q_h}&\;=\theta + \Sigma h,\\	
D_{\sigma^{Q_h}}&\;=D_{\sigma},\\
\Omega^{Q_h}&\;=\Omega.\\	
\end{split}
\end{equation}
The expression of the Esscher risk-neutral characteristic function of the $j$-th log-return process is
\begin{equation}\label{eq:CfMGHMarginalRn}
\begin{split} 
\Psi_{Y_t^{j}}\left(u\right)&=\exp\left(i u_j \mu_j^{Q_ h}\right)\left[1-\frac{2}{\psi^{Q_ h}}\left(iu_j\theta_j^{Q_ h}-\frac{1}{2} u_j^2{\sigma^{Q_ h}}^2 \right)\right]^{-\frac{\epsilon^{Q_ h}}{2}} \times\\& \frac{K_{\epsilon^{Q_ h}}\left(\sqrt{\chi^{Q_ h}\left( \psi^{Q_ h}-2 \left(iu_j\theta_j^{Q_ h}-\frac{1}{2} u_j^2{\sigma^{Q_ h}}^2\right)\right)}\right)}{K_{\epsilon^{Q_ h}}\left(\sqrt{\chi^{Q_ h} \psi^{Q_ h} }\right)}
\end{split} 
\end{equation}

Note that the Esscher change of probability measure does not modify the nature of joint and marginal log-return processes (compare (\ref{eq:ChfMGH}) with (\ref{eq:CfMGHRn})),  
and (\ref{eq:CfMGHMar}) with (\ref{eq:CfMGHMarginalRn})). Only the parameter $\psi$ and the vector $\theta$ change. In particular, the risk-neutral log-return process is a multivariate Brownian motion with correlated components, time-changed by a single generalized inverse Gaussian subordinator independent of the Brownian motion. Even if the correlation matrix of the underlying Brownian motion is not affected by the change of measure, the risk-neutral correlation matrix of log-returns is different (see (\ref{eq:MGHEsscherTransformParameters}) and (\ref{eq:CorrelationMGH})). Notice also that all marginal moments change (not only the mean). It should also be pointed out that all risk-neutral moments depend on the entire dependence structure.  As shown in equation (\ref{eq:MGHEsscherTransformParameters}), 
the risk-neutral distribution depends on the correlation matrix of the underlying Brownian motions under the historical measure.

In order to find the inverse Esscher transform, that is the vector $h$ that allows to find the historical parameters (under the historical measure $P_h$) starting from the risk-neutral ones (under the risk-neutral measure $Q$), from the system (\ref{eq:InvEsscherSystem}) it follows that the system to be solved to find $h$ is the following
\begin{equation}\label{eq:InverseMGHSystem}
\begin{cases}
{\mu_1}^{Q}-\frac{\epsilon^{Q}}{2}\ln\left[1-\frac{2}{\psi^{Q}}\left({\theta_1}^{Q}+\frac{1}{2} {\sigma_1^{Q}}^{2}\right)\right] +\ln\frac{K_{\epsilon^{Q}}\left(\sqrt{\chi^{Q}\left( \psi^{Q}-2\left({\theta_1}^{Q}+\frac{1}{2} {\sigma_1^{Q}}^{2}\right)\right)}\right)}{K_{\epsilon^{Q}}\left(\sqrt{\chi^{Q} \psi^{Q} }\right)}
=r-d_1,\\ 
  \vdots&\\
{\mu_n}^{Q}-\frac{\epsilon^{Q}}{2}\ln\left[1-\frac{2}{\psi^{Q}}\left({\theta_n}^{Q}+\frac{1}{2} {\sigma_n^{Q}}^{2}\right)\right] +\ln\frac{K_{\epsilon^{Q}}\left(\sqrt{\chi^{Q}\left( \psi^{Q}-2\left({\theta_n}^{Q}+\frac{1}{2} {\sigma_n^{Q}}^{2}\right)\right)}\right)}{K_{\epsilon^{Q}}\left(\sqrt{\chi^{Q} \psi^{Q} }\right)}
=r-d_n,\\ 
\mu^{P_ h}=\mu^{Q},\\
\epsilon^{P_ h}=\epsilon^{Q},\\	
\chi^{P_ h}=\chi^{Q},\\	
\psi^{P_ h}={\psi^{Q}}+2\left(h'\theta^Q - \frac{1}{2} h'\Sigma^{Q} h\right),\\	
\theta^{P_ h}=\theta^{Q} - \Sigma^{Q} h,\\	
D_{\sigma^{P_h}}=D_{\sigma^{Q}},\\
\Omega^{P_h}=\Omega^{Q}.\\	
\end{cases}
\end{equation}In solving this system one has to consider that, given the parameters under the measure $Q$, both the vector $h$ and the parameters under the measure $P_h$ are the system unknowns to be found. As discussed in Section \ref{sec:Model}, the system has to be solved under the constraints in equation (\ref{eq:MGHEsscherTransformParameters}) and, thus, the system unknowns are the vector $h$ together with the parameters $\psi^{P_ h}$ and $\theta^{P_ h}$.

\section{Data}\label{sec:Calibration}

In this section we provide a description of the data used in the empirical analysis. In the first empirical test we consider the same dataset analyzed in \cite{tb2014ijtaf}, that is, daily dividend adjusted closing prices from January 2, 1990 through December 31, 2012 and implied volatilities from January 2, 2008 to December 31, 2012 obtained from Bloomberg for five selected companies included in the S\&P 500: Apple Inc. (ticker APPL), Dell Inc. (ticker DELL),  International Business Machines Corp. (ticker IBM), Hewlett-Packard Comp. (ticker HPQ), Microsoft Corp. (ticker MSFT) representing five major multinational information technology companies. The implied volatilities are extracted from European call and put options with a maturity between one month and one year and with moneyness between 80\% and 120\%. That dataset is made up of more than 50,000 observations for each company. 

Then, for further empirical investigations we use the daily logarithmic return series for all Euro denominated stocks included in the EuroStoxx 50 on April 30, 2017.\footnote{ We select Unicredit Bank and Assicurazioni Generali instead of CRH and Engie.} We obtained from Datastream daily dividend-adjusted closing prices from June 30, 2002 through April 30, 2017. Furthermore, implied volatilities were extracted from European call and put options written on selected stocks from June 30, 2009 to April 30, 2017 with a one month maturity and with moneyness between 80\% and 120\%. As risk-free interest rate we take the Euribor rate. Since we considered dividend adjusted closing prices, we assumed that $d_j=0$ for each stock $j$. By an empirical test it follows that under this assumption on dividends the put-call parity continues to be fulfilled.

\section{Double calibration}\label{sec:TripleCalibration}

The failure to explain observed option prices by considering only time series information is well known (see \cite{chernov2000study}). For this reason, in this Section we consider a calibration framework in which we jointly estimate the model parameters on the time series of log-returns, by minimizing  (1) the Kolmogorov-Smirnov distance under the historical measure, and calibrate the implied volatility surface, by (2) minimizing the average relative percentage error (ARPE), under the risk-neutral measure.  We refer to this method as {\it double calibration} (\cite{tb2014ijtaf}). By following this approach both the implied volatility and log-return calibration-estimation errors are minimized.

From a practical perspective, on each trading day we solve the following minimization problem
\begin{equation}\label{eq:OptimizationProblem}
\widehat{\Theta}^Q = \min_{\Theta^Q}\Big( \sum_j\big( ARPE_j(\Theta^Q) +  \xi_1 KS_j(\Theta^{P_h})\big)\Big),
\end{equation}
where the ARPE is given by 
\begin{equation}\label{eq:ARPE}
ARPE_j(\Theta^Q) = \frac{1}{\text{number of observations}}\sum_{T_n}\sum_{K_m}\frac{|iVol^{market}_{T_n K_m} - iVol^{model}_{T_n K_m}(\Theta^Q)|}{iVol^{market}_{T_n K_m}},
\end{equation} 
where $iVol^{market}_{T_n K_m}$ ($iVol^{model}_{T_n K_m}$) denotes the market (model) implied volatility of the option with maturity $T_n$ and strike $K_m$, the index $j$ represents the stock $j$, and $\Theta^Q$ is the parameter vector according to a given model under the risk-neutral measure $Q$. Then, $KS_j$ is the Kolmogorov-Smirnov distance of the margin $j$ given the set of parameters $\Theta^{P_h}$ computed through the inverse Esscher transform, that is the measure change that allows to compute the historical parameters  $\Theta^{P_h}$ under the measure $P_h$ given the risk-neutral parameters $\Theta^Q$ under $\Q$. To find this inverse transform, we solve the system (\ref{eq:InverseMGHSystem}) and (\ref{eq:InverseMNTSSystem}) in the MGH and in the MNTS model, respectively. After some attempts, we fixed $\xi_1$ equal to 3. This value for $\xi_1$ shows a good balance between model performance and parameter stability. 

In practice, we want to find a set of parameters $\Theta^Q$ such that the model implied volatility ($iVol^{model}$) is as close as possible to the market implied volatility ($iVol^{market}$) and, at the same time, the theoretical univariate distribution of log-returns is as close as possible to the empirical distribution. Since the minimization problem (\ref{eq:OptimizationProblem}) with respect to the parameter vector $\Theta^Q$ has not a closed-form solution and it may not have a global minimum, a numerical optimization routine is needed to find a local minimum. We use the Matlab r2016b function {\it fmincon} for the optimization routine and a function found on the Paul Wilmott web-site to compute the implied volatilities from the values of option prices. We follow the analytical (up to an integration) pricing method for standard vanilla options proposed in \cite{carr1999option} and the expectation-maximization (EM) maximum likelihood estimation method (see \cite{tb2014ijtaf} and \cite{btf2015riding}) to 
find a proper starting point of the optimization procedure.

The minimization problem (\ref{eq:OptimizationProblem}) is ill-posed, mainly because the solution is not necessarily unique and there is no guarantee that a solution exists. However, the numerical procedure we adopted shows satisfactory results. More precisely, we use the following procedure:
\begin{enumerate}
\item on the first observation day $t=1$, we perform the EM-based maximum likelihood estimation on the time series of log-returns, and estimate the set of historical parameters $\Theta^{P}$ under the measure $\P$;
\item on the first observation day $t=1$, we compute the Esscher transform, that is, given the set of historical parameters  $\Theta^{P}$, we compute the corresponding risk-neutral parameters $\Theta^{Q_h}$;
\item the parameters computed on the previous step are needed as starting point of the numerical procedure used to find a solution of the optimization problem (\ref{eq:OptimizationProblem}); 
\item we run a function that performs the following computations:
\begin{itemize}
 \item it computes the implied volatilities and the corresponding ARPE for each stock;
 \item by solving the system (\ref{eq:InverseMGHSystem}) and (\ref{eq:InverseMNTSSystem}) in the MGH and in the MNTS model, respectively, it computes the inverse Esscher transform, that is, given the set of risk-neutral parameters $\Theta^Q$, it finds the corresponding historical parameters $\Theta^{P_h}$ (the Matlab {\it fsolve} function has been considered to find a solution for the systems (\ref{eq:InverseMGHSystem}) and (\ref{eq:InverseMNTSSystem}));
 \item by considering the historical parameters $\Theta^{P_h}$, it computes the KS distance under the historical measure $\P$ for each stock; 
\end{itemize}
thus, we cast the function into the algorithm that finds a solution for the optimization problem (\ref{eq:OptimizationProblem});
\item we move to day $t + 1$ and find the matrix $A$ through the EM-based maximum likelihood estimation on the time series of log-returns;
\item we return to Step 4 and choose as starting point of the optimization problem the solution found on the day $t$ and the matrix $A$ estimated on the previous Step 5.
\end{enumerate}

We point out that the matrix $A$ estimated through the EM-based maximum likelihood method is kept fixed into the optimization procedure. This in practice means that all other parameters may change into the optimization algorithm except the matrix $A$ representing the lower triangular Cholesky factor of the correlation matrix $\Omega$. From equations (\ref{eq:MNTSEsscherTransformParameters}) and (\ref{eq:MGHEsscherTransformParameters}) it follows that $\Sigma^Q=\Sigma^{P_h}$. It is noted that the matrix $\Omega$ represents the correlation matrix of the underlying Brownian motions and the correlation matrix of the random vector $Y_t$ is given by equations (\ref{eq:CorrelationMNTS}) and (\ref{eq:CorrelationMGH}). Furthermore, we remind that the matrix $\Omega$ is not affected by the change of measure.   

\subsection{Comparison with the MNIG and the MVG model}\label{sec:Comparison}

\begin{figure}
\begin{center}
\includegraphics[width=1\columnwidth]{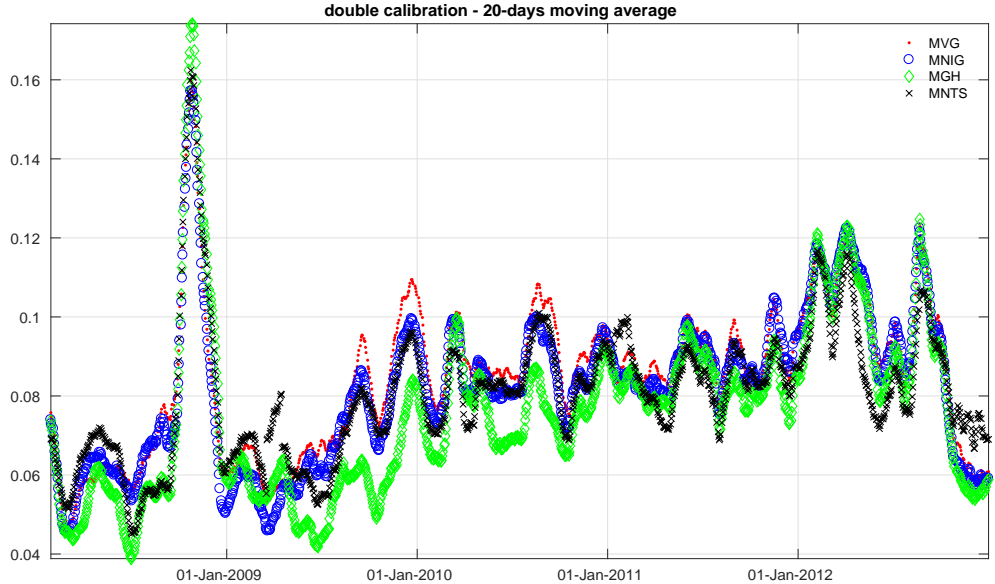}
\caption[{\it Calibration error time series}]{\label{fig:ARPEtimeseries}\footnotesize Implied volatility calibration error (ARPE) for all stocks and models analyzed under the double calibration approach (20-day moving average). The calibration was conducted on a daily basis for each trading day between January 2, 2008 to December 31, 2012.}
\end{center}
\end{figure}

In this section we compare the calibration errors of both the MGH and the MNTS model with those reported in \cite{tb2014ijtaf}, that is with the multivariate Gaussian, the multivariate variance gamma (MVG) and the multivariate normal inverse Gaussian (MNIG) model. Based on the ARPE evaluated over the entire sample on successive cross-sections of implied volatilities and stocks, the MGH model shows a smaller calibration error in fitting implied volatilities. The error is larger for IBM (in median, 9.25 per cent in MGH model, and 10.20 per cent in the MNTS model) and smaller for Dell (in median, 6.40 per cent in the MGH model, and 6.50 per cent in the MNTS model). The time series of the ARPE computed across all five stocks simultaneously ranges from 3.16 per cent to 21.38 per cent (in median, 7.18 per cent) for the MGH model, and from 3.34 per cent to 32.71 per cent (in median, 7.74 per cent) for the MNTS model. 

On October 17, 2008 (March 19, 2009) the MGH (MNTS) model reaches the largest calibration error. As already observed in \cite{guillaume2012sato}, multivariate models based on L\'evy processes performed badly during the crisis period. In Figure \ref{fig:ARPEtimeseries} we report the time series of the 20-day moving average of the median ARPE computed across all five stocks, for both the MGH and the MNTS model, and compare them with the MNIG and MVG models analyzed in \cite{tb2014ijtaf}. For each stock and for each model we evaluate the ARPE over the entire period. The 20-day moving average ranges from 3.88 per cent and 17.42 per cent (on average, 7.61 per cent) in the MGH case, from 4.51 per cent and 16.24 per cent (on average, 8.05 per cent) in the MNTS case, from 4.54 per cent and 15.72 per cent (on average, 8.18 per cent) in the MNIG case, and from 4.71 per cent and 15.97 per cent (on average, 8.47 per cent) in the MVG case. As expected the MGH model has a smaller average implied volatility calibration error than both the MNIG and the MVG model. 
The MNTS model has also a slightly better performance compared with both the MNIG and the MVG model. Additionally, by considering similar studies on this subject (see \cite{bianchi2013smile}) the implied volatility calibration error, for both the MGH and the MNTS model, is satisfactory.

\begin{figure}
\begin{center}
\includegraphics[width=1\columnwidth]{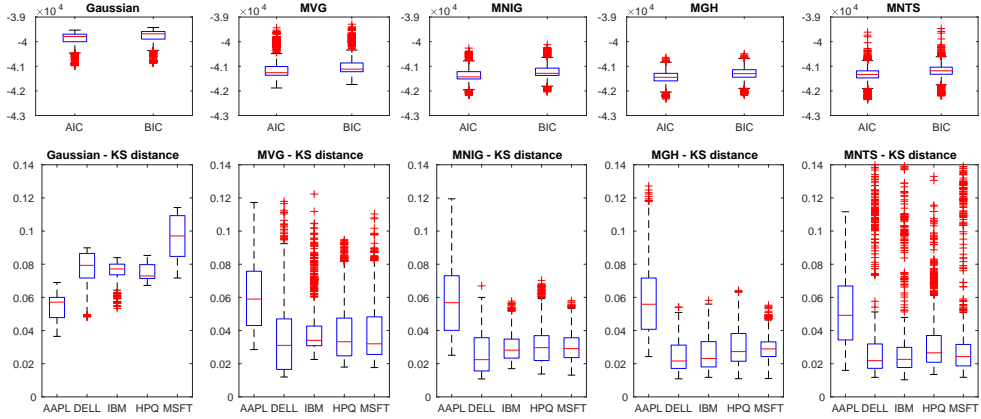}
\caption[{\it Goodness of fit}]{\label{fig:subplotsTests}\footnotesize Goodness of fit statistics of the models estimated using the double calibration approach for each trading day from January 2, 2008 to December 31, 2012. For each trading day, a window of fixed size is considered (1,500 trading days) for a total of 1,259 rolling windows estimations for each model. For each model the boxplots of the AIC, of the BIC and of the univariate KS distance are reported. On each boxplot, the central mark is the median, the edges of the box are the 25-$th$ and 75-$th$ percentiles, the whiskers extend to the most extreme data points not considered outliers, and outliers are plotted individually.}
\end{center}
\end{figure}

After having analyzed the implied volatility calibration error, we describe the performance in fitting the behaviour of time series of log-returns. We perform this analysis for both the MGH and the MNTS model and compare them with the MNIG and MVG models analyzed in \cite{tb2014ijtaf}. For all calibrated models, in Figure \ref{fig:subplotsTests} we report the boxplot of both the Akaike information criterion (AIC) and the Bayesian information criterion (BIC). According to both the AIC and the BIC, the MGH  is the best performing model and the Gaussian is the worst one. We point out that we conducted a maximum likelihood estimation for Gaussian model, without computing the corresponding implied volatility calibration error. 
Additionally, Figure \ref{fig:subplotsTests} shows the boxplots of the KS statistic for each margin. The average $p$-value ranges from 2.34 (APPL) to 45.79 (DELL) per cent in the MGH model, and from 5.51 (APPL) and 43.46 (DELL) per cent in the MNTS model. 

We remind that the historical parameters $\Theta^{P_h}$ are computed through the inverse Esscher transform. Except for Apple and, particularly for the MNTS model, the KS statistics are smaller compared to the KS statistics computed under the multivariate normal assumption. As shown in Figure \ref{fig:subplotsTests} both the MNIG and the MVG model show a worst performance than both the MGH and MNTS model. The average $p$-value ranges from 1.59 (APPL) to 45.19 (DELL) per cent in the MNIG model, and from 0.88 (APPL) and 35.48 (DELL) per cent in the MVG model. While the MGH model outperforms its competitor models in calibrating the volatility surface, the MNTS model is slightly better in explaining the behaviour of time series of log-returns, at least for the data considered in this section. In Section \ref{sec:Portfolio} we further analyze these models in a portfolio allocation exercise.

Finally, to perform a double calibration on a given trading day, that is to calibrate the five observed volatility surfaces and simultaneously estimate the five time series of log-returns, the computing time is, in median, 90 seconds in the MNIG case, and 190 seconds in the MVG case. The computing time for both the MGH and the MNTS model is larger (respectively, 200 and 400 seconds). This procedure was run on an 8 cores AMD FX 8120 processor with 16GB of Ram with a Linux based 64-bit operating system. Note that the optimization function makes use of the Matlab Parallel toolbox. Most of the computing time is spent on the optimization part of the code. 

In the optimization algorithm we constrain the three MGH parameters ($\lambda$, $\chi$, $\psi$) in the region between (-4.5, 1e-2, 1e-2) and (-0.5, 5, 2), the three MNTS parameters ($\omega$, $\lambda$, $C$) in the region between (0.75, 1e-2, 1e-2) and (1.75, 5, 100). While in the MGH (MNTS) case the parameter  $\sigma_{j}$ ranges between 0.01  and 0.15 (0.01 and 0.2), $\theta_j$ ranges between -0.1 and 0.01  (-0.15 and 0.01). The optimization algorithm applied in this study is a sequential quadratic programming method implemented in the {\it fmincon} Matlab function in which the option {\it active-set} is selected with the \textit{UseParallel} option always switched on.

\subsection{A large scale empirical test}\label{sec:EurostoxxCalibration}

\begin{figure}
\begin{center}
\includegraphics[width=1\columnwidth]{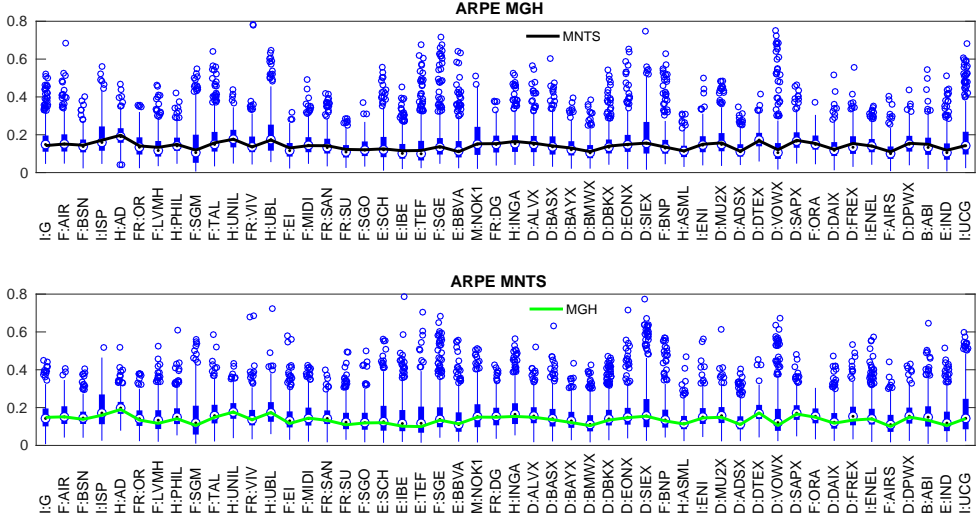}
\caption[{\it Calibration error time series}]{\label{fig:ARPEboxplotEurostoxxWeekly}\footnotesize Implied volatility calibration error (ARPE) for each stock and each model analyzed under the double calibration approach (boxplot). For each stock the boxplot of the ARPE of a given model is compared to the median values of the ARPE of the competitor models. The calibration was conducted on a weekly basis for each Wednesday between June 30, 2009 to April 31, 2017.}
\end{center}
\end{figure}

\begin{figure}
\begin{center}
\includegraphics[width=1\columnwidth]{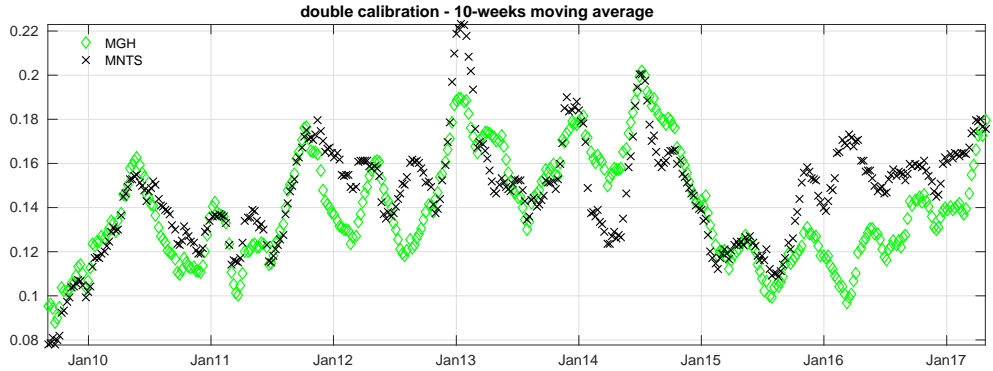}
\caption[{\it Calibration error time series}]{\label{fig:ARPEtimeseriesEurostoxxWeekly}\footnotesize Implied volatility calibration error (ARPE) for all stocks and models analyzed under the double calibration approach (10-weeks moving average). The calibration was conducted on a weekly basis for each Wednesday between June 30, 2009 to April 31, 2017.}
\end{center}
\end{figure}

For further empirical investigation, we apply the calibration described in Section \ref{sec:TripleCalibration} to a large scale case. That is, we assess the double calibration approach to fit both the MGH and the MNTS model on all Euro denominated stocks included in the EuroStoxx 50 and we compare them with the multivariate Gaussian model with parameters $\mu$ and $\Sigma$. As usual, $\mu$ represents the annualized mean vector and $\Sigma$ is the annualized variance-covariance matrix. Both parameters are calibrated only to the time series of stock log-returns. We point out that the two non-Gaussian models have 53 more parameters compared to the Gaussian model, that is, the three parameters of the subordinator and the 50-dimensional vector $\theta$.

The calibration is conducted for each Wednesday from June 30, 2009 to April 31, 2017. For each Wednesday, a window of fixed size is considered (1,500 daily observations) for a total of 409 rolling windows estimations for each model. The models are fitted on these time series of stock log-returns and to the 50 one-month implied volatility smiles observed at each given Wednesday.

\begin{figure}
\begin{center}
\includegraphics[width=1\columnwidth]{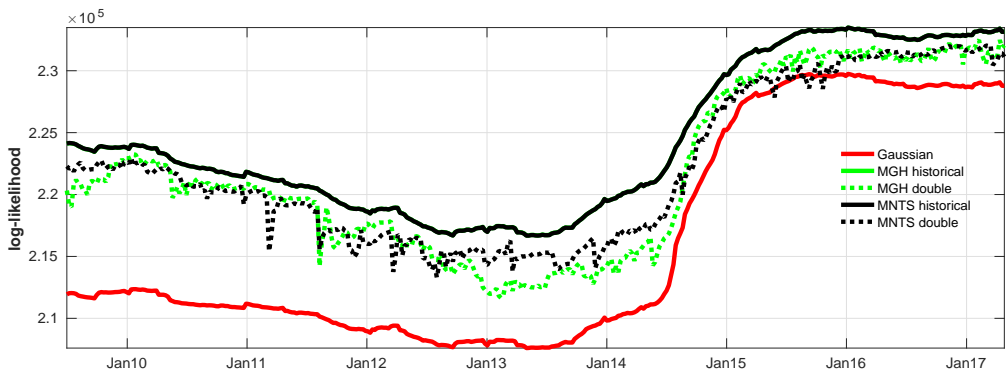}
\caption[{\it Log-likelihood}]{\label{fig:LLtimeseriesEurostoxxWeekly}\footnotesize Log-likelihood of the models estimated using the historical and the double calibration approach for each Wednesday from June 30, 2009 to April 31, 2017. For each trading day, a window of fixed size is considered (1,500 trading days) for a total of 409 rolling windows estimations for each model.}
\end{center}
\end{figure}

\begin{figure}
\begin{center}
\includegraphics[width=1\columnwidth]{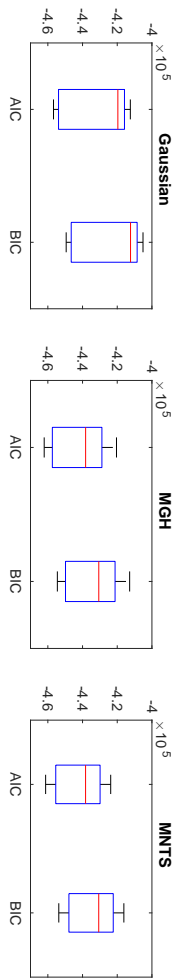}
\includegraphics[width=1\columnwidth]{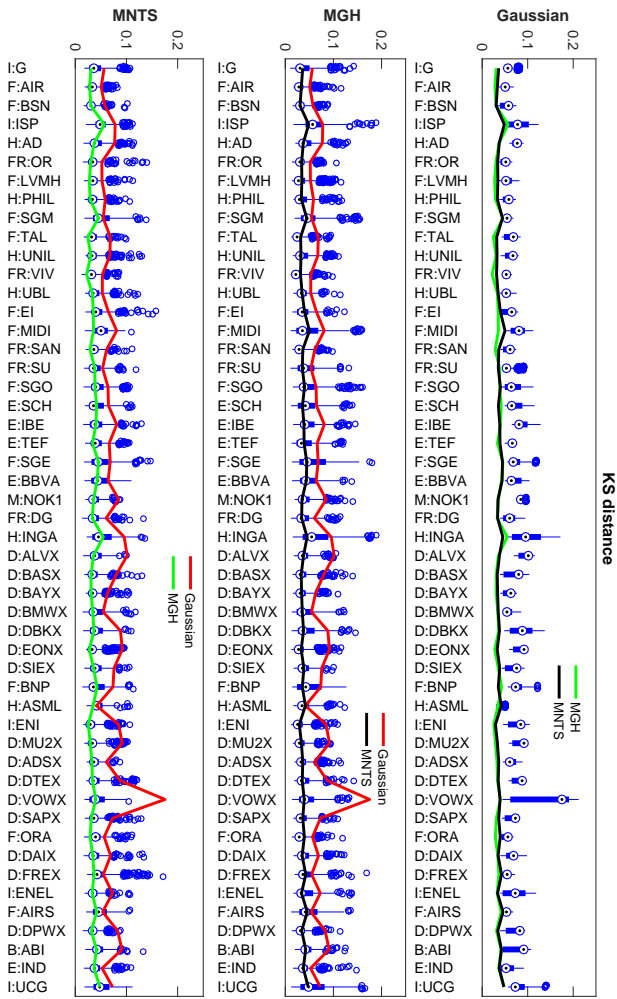}
\caption[{\it Goodness of fit}]{\label{fig:subplotsTestsEurostoxxWeekly}\footnotesize Goodness of fit statistics of the models estimated using the double calibration approach for each Wednesday from June 30, 2009 to April 31, 2017. For each trading day, a window of fixed size is considered (1,500 trading days) for a total of 409 rolling windows estimations for each model. For each model the boxplots of the AIC, of the BIC and of the univariate KS distance are reported. For each stock the boxplot of the KS distance of a given model is compared to the median values of the KS distance of the competitor models. On each boxplot, the central mark is the median, the edges of the box are the 25-$th$ and 75-$th$ percentiles, the vertical line extend to the most extreme data points not considered outliers, and outliers are plotted individually.}
\end{center}
\end{figure}

Based on the ARPE evaluated over the entire sample of implied volatilities and stocks, the MGH model shows a smaller calibration error in fitting implied volatilities, as shown in Figure \ref{fig:ARPEboxplotEurostoxxWeekly}, where the boxplots of the ARPE computed across the 409 rolling windows are reported. While in the MGH case the median error ranges from 9.95 (Telefonica - E:TEF) to 19.03 (Ahold Delhaize - H:AD) per cent, in the MNTS case it ranges from 10.99 (Airbus - F:AIRS) to 19.76 (Ahold Delhaize - H:AD) per cent.  In 8 cases over 50 the median calibration error of the MNTS model is smaller compared to the error of the MGH model.

As shown in Figure \ref{fig:ARPEtimeseriesEurostoxxWeekly}, the time series (10-weeks moving average) of the median ARPE computed across all 50 stocks ranges from 8.80 per cent to 20.20 per cent (on average, 13.94 per cent) for the MGH model, and from 7.77 per cent to 22.30 per cent (on average, 14.64 per cent) for the MNTS model. 

In Figure \ref{fig:LLtimeseriesEurostoxxWeekly} we report the time series of the log-likelihood for all competitor models and estimation methods. The maximum likelihood estimation is considered in the normal case, the EM-based maximum likelihood estimation method in the MGH historical and MNTS historical cases, the double calibration approach is applied in the MGH double and MNTS double cases.  Even if the double calibration approach is not focused on maximizing the likelihood, there are only few days when the log-likelihood of the MNTS model is smaller that the likelihood of the normal model (the log-likelihood of the MGH model is never smaller). This shows the great flexibility of the MGH and MNTS models with respect to the normal one: they are able to jointly calibrated log-returns and implied volatility smile and their log-likelihood is still greater than the estimated log-likelihood under the normal framework. As expected, in both MGH and MNTS cases, the log-likelihood of the models estimated by considering only historical information is 
larger. The log-likelihoods of MGH historical and MNTS historical are almost indistinguishable.

As shown in Figure \ref{fig:subplotsTestsEurostoxxWeekly}, according to both the AIC and the BIC, the MNTS model is better because its AIC and BIC average values are smaller compared with all other competitor models, and the Gaussian one is the worst. Additionally, Figure \ref{fig:subplotsTestsEurostoxxWeekly} shows the boxplots of the KS statistic for each margin. In 18 cases over 50 the calibration error (average values) of the MNTS model is smaller compared to the error of the MGH one. For all stocks for both the MGH and the MNTS model the KS statistic is smaller compared to that of the Gaussian model. The $p$-value (average value) is bigger than 5 per cent in 44 cases over 50 for the MNTS model, in 49 cases for the MGH model and never for the Gaussian one. 

Finally, to perform a double calibration on a given Wednesday, that is to calibrate the 50 observed volatility surfaces, and simultaneously estimate the time series of log-returns the computing time is, in median, 350 seconds in the MGH case, and 1,350 seconds in the MNTS case. The procedure is run on the same machine and with the same implementation described in Section \ref{sec:Comparison}. 

\section{A portfolio selection analysis}\label{sec:Portfolio}

In this section we provide the necessary definitions needed to implement a minimum-AVaR portfolio selection criterion and show the backtest of a portfolio selection strategy applied to the 50-dimensional case investigated in Section \ref{sec:EurostoxxCalibration}. 

The value at risk (VaR) of a continuous random variable $X$ at tail probability level $\delta$ is 
$$
VaR_\delta (X) = -\inf\{x| P (X\leq x) > \delta\} = - F_X^{-1} (\delta)
$$
and it can be computed by inverting the cumulative distribution function $F_X$. The AVaR of a continuous random variable $X$ with finite mean (i.e. $E[X]<\infty$) at tail level $\delta$ is the average of the VaRs that are greater than the VaR at tail level $\delta$, that is
$$
AVaR_\delta (X) = \frac{1}{\delta}\int_0^{\delta} VaR_p(X) dp = - E \big[X\big|X < - VaR_{\delta}(X)\big].
$$ 
Thus, the AVaR is measures the expected loss, given that the loss has exceeded the VaR at the same probability level. We refer to $VaR_{0.05}(X)$ and $AVaR_{0.05}(X)$ as {\it 5\% VaR} and {\it 5\% AVaR}, respectively.

\begin{figure}
\begin{center}
\includegraphics[width=1\columnwidth]{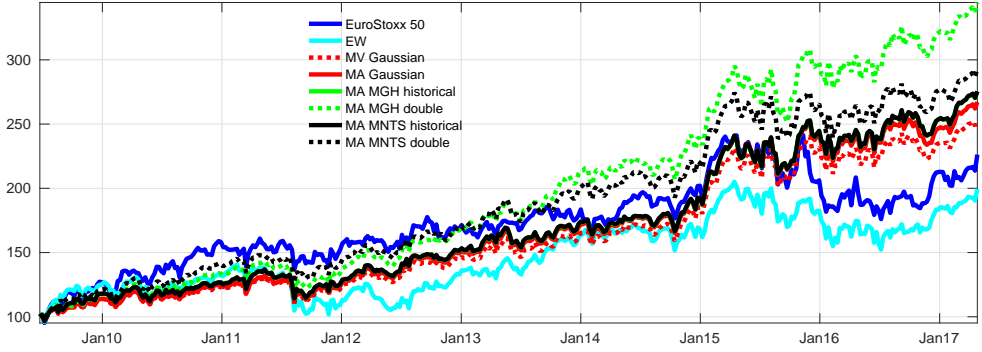}
\caption[{\it Portfolio total wealth}]{\label{fig:portfolioTotalWealth}\footnotesize Portfolio allocation backtest with weekly rebalacing for each Wednesday from June 30, 2009 to April 31, 2017 (for a total of 409 rebalancing days) for the MV, EW and MA strategies. The resulting portfolio values are scaled to 100 for the first date of the backtest period. For the MGH and the MNTS both the historical and the double calibration approaches are considered to find the MA weights.}
\end{center}
\end{figure}

By considering the result of Proposition 1 in \cite{kim2012measuring}, it follows that if $Y_t$ is a MNTS process, $Y_{\Delta t}$ is the distribution of its increments with discrete time step $\Delta t$, and $w\in\R^n$, then $w'Y_{\Delta t}$ is a NTS random variable with characteristic function as in equation (\ref{eq:CfMTSMar}) with parameters ($\omega$, $\lambda$, $C$, $\tilde{\theta}$, $\tilde{\mu}$, $\tilde{\sigma}$), where
$$
\tilde{\theta} = w'\theta \qquad \tilde{\mu} = w'\mu \qquad  \tilde{\sigma} = \sqrt{w'\Sigma w}.
$$
This means that a portfolio of MNTS margins is a NTS random variable. This property is very useful for computing portfolio risk measures, since it reduces the dimension of the problem from $n$ to 1 and the portfolio distribution belongs to the same parametric family. Furthermore, from \cite{krbf2010pms} and \cite{krbf2011jbf} it is possible to obtain a closed formula (up to an integration) to compute the average value at risk (AVaR) in the NTS case. In the MGH case, the VaR and AVaR computation is simpler, since a portfolio of MGH margins is a GH random variable and a closed-form expression for the density function can be used. In the normal case, the VaR and the AVaR can be easily evaluated. 

\begin{figure}
\begin{center}
\includegraphics[width=1\columnwidth]{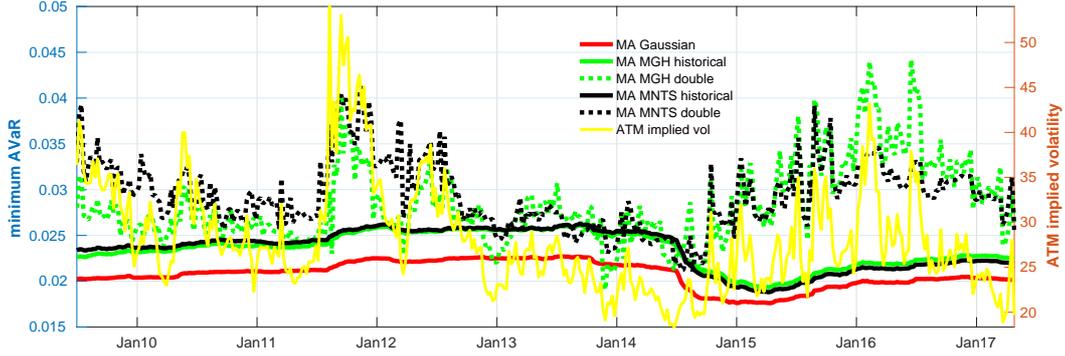}
\caption[{\it Portfolio total wealth}]{\label{fig:portfolioOptimalAVaR}\footnotesize AVaR estimates for the MA portfolio strategy with weekly rebalacing for each Wednesday from June 30, 2009 to April 31, 2017 (for a total of 409 rebalancing days). For the MGH and the MNTS both the historical and the double calibration approach are considered to find the MA weights.}
\end{center}
\end{figure}

The portfolio selection strategy proposed here is based on the minimum-AVaR (MA) approach. Given a distributional assumption, we find the weights minimizing the AVaR (see \cite{stoyanov2010survey}) at a given tail level. We suppose that
short selling is not allowed,  that is not possible to invest more than 10\% of the wealth in a specific stock
($0\leq w_j \leq 0.1$) and we consider $\delta=0.05$. Due to the convexity property of AVaR, the problem has a unique minimum which can be obtained through the standard first-order optimality conditions for constrained optimization problems. We apply this approach to the Gaussian, the MGH and the MNTS models. In the non-Gaussian cases we consider both the historical and the double calibration approach. The former calibration is based on the time-series of stock log-returns only, for this reason we refer to it as {\it historical}. As described in Section \ref{sec:TripleCalibration}, the latter uses both time-series of stock log-returns and option implied volatility, and we refer to it as {\it double}.

The performance of the proposed long-only strategy is benchmarked against the minimum-variance (MV) portfolio and the equally weighted (EW) portfolio with rebalancing. These fundamental strategies are important benchmarks for large-scale applications. As done in \cite{mainik2015portfolio}, the comparison includes annualized portfolio returns, maximum drawdowns, transaction costs, portfolio concentration, and asset diversity in the portfolio. Note that none of the three compared methods (MV, EW, and MA) looks at expected returns.

The analysis is based on the estimates described in Section \ref{sec:EurostoxxCalibration}. The computation of portfolio weights utilizes the estimates based on the time series of stock log-returns from the six years prior to each Wednesday and the one-month implied volatility smile observed on that Wednesday. For example, the optimal portfolio for July 1, 2009 is estimated from the stock price data for the period from July 2, 2003 to July 1, 2009 and the one-month implied volatilities observed on July 1, 2009. While the estimates are based on daily data, the rebalancing is performed on a weekly basis. Estimating both the MGH and the MNTS model on each trading day is too time-consuming for our computing resources. As observed in Section \ref{sec:EurostoxxCalibration}, for each estimation day, it needs around half an hour to calibrate the two non-Gaussian models (MGH and MNTS). 

In Figure \ref{fig:portfolioTotalWealth} we report the behavior of the total wealth for all strategies and for the EuroStoxx 50 index. The estimated optimal AVaR of the MA strategy applied to different distributional assumptions (Gaussian, MGH and MNTS) and estimation methods (historical and double) is reported in Figure \ref{fig:portfolioOptimalAVaR}. The dynamics of the AVaR are compared with the at-the-money (ATM) implied volatility. By construction the optimal AVaR obtained by considering MGH and MNTS estimates based on the double calibration approach is more volatile. Note that the double calibration approach takes into consideration the behavior of the implied volatility, that is usually less smooth than the historical one. 

In Table \ref{tab:portfolioBacktest} we show the results of the MA approach compared to its competitor strategies and to the EuroStoxx 50 index. We report the total return (TR) over the observation period, the corresponding annualized return (AR), the Sharpe ratio, the maximum drawdown (MaxDD), the concentration coefficient (CC) and the portfolio turnover (PT).

The Sharpe ratio is estimated on the portfolio returns over the observation period (409 weeks). To measure the portfolio stock concentration, the concentration coefficient (CC), also known as Herfindahl-Hirschman index, defined as
$$
CC_t = \left(\sum_{j=1}^n \left(w_t^j\right)^2\right)^{-1}
$$
is computed, where $w_t^j$ is the portion of portfolio wealth invested in the $j$-th stock at time $t$. The CC of an equally weighted portfolio is the number of assets $n$. As the portfolio becomes concentrated on fewer assets, the CC decreases. 

\begin{table}
\begin{center}
\begin{footnotesize}
\begin{tabular}{@{}lcccccc@{}}
\toprule
	&	TR	&	AR 	&	Sharpe	&	MaxDD	&	 CC 	&	 PT 	\\

EuroStoxx 50	&	81.6\%	&	10.4\%	&	7.2\%	&	27.1\%	&	 -   	 & 	 -   	\\
EW	&	69.1\%	&	8.8\%	&	5.9\%	&	27.7\%	&	 50.0 	 & 	 0.0202 	\\
MV normal	&	93.0\%	&	11.8\%	&	10.7\%	&	15.8\%	&	 12.3 	 & 	 0.0161 	\\
MA normal	&	98.3\%	&	12.5\%	&	11.3\%	&	15.2\%	&	 12.4 	 & 	 0.0160 	\\
MA MGH historical	&	101.3\%	&	12.9\%	&	11.6\%	&	16.6\%	&	 12.1 	 & 	 0.0157 	\\
MA MGH double	&	123.7\%	&	15.7\%	&	14.1\%	&	16.1\%	&	 11.8 	 & 	 0.0162 	\\
MA MNTS historical	&	101.3\%	&	12.9\%	&	11.6\%	&	16.6\%	&	 12.2 	 & 	 0.0157 	\\
MA MNTS double	&	107.5\%	&	13.7\%	&	12.3\%	&	14.5\%	&	 11.8 	 & 	 0.0158 	\\

\bottomrule
\end{tabular}
\caption[{\it Portfolio backtest}]{\label{tab:portfolioBacktest}\footnotesize Portfolio allocation backtest with weekly rebalacing for each Wednesday from June 30, 2009 to April 31, 2017 (for a total of 409 rebalancing days) for the MV, EW and MA strategies. We consider the following performance measures: total return (TR), annualized return (AR), Sharpe ratio (Sharpe), maximal drawdown (MaxDD), concentration index (CC) and portfolio turnover (PT). The AVaR is computed by considering a $5\%$ tail level.}
\end{footnotesize}
\end{center}
\end{table}

As a proxy for transaction costs, we consider the portfolio turnover (PT) defined as
$$
PT_t = \sum_{j=1}^n|w_t^j - w_{t_-}^j|
$$
where $w_t^j$ is the portfolio weight of the asset $j$ after rebalancing (according to the portfolio allocation strategy) at time $t$, and $w_{t_-}^j$ is the portfolio weight of the asset $j$ just before rebalancing.   

The results show that the MA strategy indeed outperforms MV and EW portfolios in many respects. In particular, the MA optimal portfolio gives higher total returns, higher Sharpe ratios, and lower maximal drawdowns. Furthermore, the use of the information content of the implied volatility largely improve the portfolio performance. The strategies where the parameters are estimated with the double calibration approach outperforms all competitor strategies. 
The best performer is the MA MGH double portfolio with the highest returns (15.7 per cent on an annual basis) and the highest Sharpe ratio (14.1 per cent). The overperformance on a yearly basis with respect to the EuroStoxx 50 index is 5.3 (3.3) per cent in the MA MGH (MNTS) double case, 2.5 per cent in the MA MNTS (MGH) historical case, 2.1 per cent in the MA Gaussian case and 1.4 in the MV Gaussian case. The EW strategy underperforms the EuroStoxx 50 index. The lowest maximal drawdown is obtained for the MA MNTS strategy (14.5 per cent), even if the value obtained for the MA normal strategy is quite close (15.2 per cent). It is slightly higher in the other non-Gaussian MA cases: it ranges from 16.1 to 16.6 per cent.

The results in Table \ref{tab:portfolioBacktest} indicate that the MA and the MV strategies are quite selective, whereas the number of stocks in the MA portfolio under the double calibration approach is slightly smaller. The average turnover of MA optimal portfolios ranges from 0.0157 to 0.162  and it is close to that of the MV portfolio (0.0161). Surprisingly the EW strategy shows the highest PT (0.0202).

\begin{figure}
\begin{center}
\includegraphics[width=1\columnwidth]{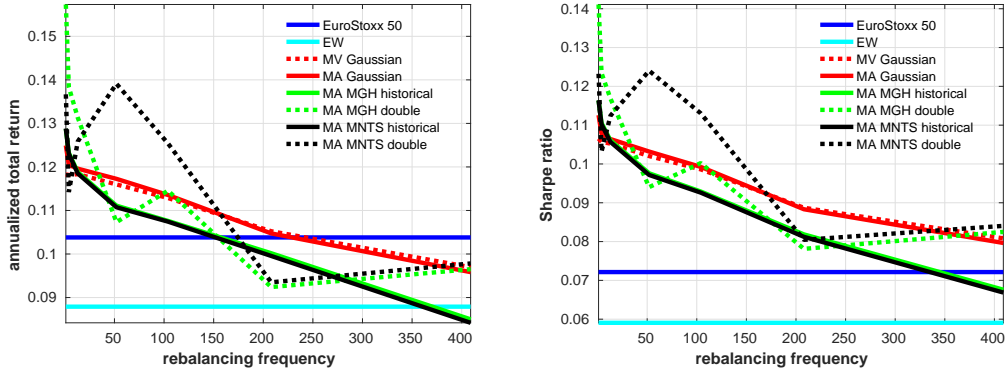}
\caption[{\it Portfolio rebalancing}]{\label{fig:portfolioRebalancing}\footnotesize For each model we compare the portfolio strategy performance by switching from weekly to monthly (four weeks), to quarterly (13 weeks), to semi-annually (26 weeks), to annually (52 weeks), to biennially (104 weeks), to quadrennially (208 weeks) rebalancing and to the buy-and-hold (no rebalacing over the entire observation period). Total annualized returns and Sharpe ratios are reported.}
\end{center}
\end{figure}

As a further empirical investigation, we switch from weekly to monthly (four weeks), to quarterly (13 weeks), to semi-annually (26 weeks), to annually (52 weeks), to biennially (104 weeks), to quadrennially (208 weeks) rebalancing and to the buy-and-hold (no rebalacing over the entire observation period). The calculation of portfolio weights is still based on the estimates provided in Section \ref{sec:EurostoxxCalibration}. This allows using all observations in the historical window, and not only a subset. The Sharpe ratio is estimated on the portfolio returns over the observation period (409 weeks). Both the AR and the Sharpe ratios are reported in Figure \ref{fig:portfolioRebalancing}. While for the strategies based only on historical information, the performance decreases if one decreases the rebalacing frequency, for both strategies that use also the implied volatility information, the performance seems to be less affected by the rebalancing frequency.

\section{Conclusion}\label{sec:Conclusions}

The objective of this paper is threefold. First, we propose a multivariate option pricing framework based on heavy tails, negative skewness and asymmetric dependence. The connection between the historical measure and the risk-neutral measure is given by the Esscher transform. This link allows one to take into account simultaneously both multivariate time-series of log-returns and implied volatility smiles.

Second, we conduct a large scale empirical study based on a joint calibration-estimation of the univariate option surfaces and of the time series of log-returns has been proposed. The model is calibrated without the need of multivariate derivative quotes. The EM-based maximum likelihood estimation method is applied to have a first estimate of the historical parameters. Thus, we jointly estimate the model parameters on the time series of log-returns by minimizing (1) the average relative percentage error, which is a measure of the distance between model and observed implied volatilities, and (2) the Kolmogorov-Smirnov distance between the theoretical and empirical historical distributions. The historical measure and the risk-neutral one are connected through the Esscher transform.

Third, we show how to use the proposed framework to evaluate portfolio risk measures. The models analyzed allow for a quasi-closed form solution for the evaluation of both VaR and AVaR. Forecasts of risk measures for portfolios of assets can be obtained in a computationally straightforward manner and the model can be cast into a portfolio optimization algorithm to efficiently solve a portfolio selection problem. We empirically assess the importance of considering the information coming from implied volatility smiles under a minimum-risk portfolio allocation strategy with multivariate non-Gaussian distributions. 

Finally, the multivariate non-Gaussian models proposed in this work together with the double calibration approach can be used to explore the interdependencies in financial markets, not only for solving portfolio allocation problems, but also as a statistical tool for financial stability purposes. This tool for dependence modeling does not only allow for an accurate analysis beyond the linear correlation matrix and the common multivariate Gaussian distribution, but it is also able to incorporate market expectations through the use of option implied volatilities.

\section*{Acknowledgments}

This publication should not be reported as representing the views of the Banca d'Italia. The views expressed are those of the authors and do not necessarily reflect those of the Banca d'Italia.

\end{document}